\documentclass[12pt,a4paper,final]{iopart}

\usepackage{graphicx}
\usepackage[breaklinks=true,colorlinks=true,linkcolor=blue,urlcolor=blue,citecolor=blue]{hyperref}
\usepackage{lipsum}
\usepackage{nicefrac}
\usepackage{color}
\usepackage{hyperref}
\usepackage{float}
\usepackage{txfonts}
\begin{document}

\title[Pancharatnam phase in non-separable states of light ]{Pancharatnam phase in non-separable states of light}
\author{Chithrabhanu Perumangatt$^{1}$}
\address{$^1$Physical Research laboratory, Navarangpura, Ahmedabad, India - 380009}
\ead{chithrabhanu@prl.res.in}

\author{Salla Gangi Reddy}
\address{$^1$Physical Research laboratory, Navarangpura, Ahmedabad, India - 380009}
\address{$^2$University of Electro-communication, Chofu, Tokyo, Japan - 1828585}
\ead{sgreddy@prl.res.in}

\author{Nijil Lal}
\address{$^1$Physical Research laboratory, Navarangpura, Ahmedabad, India - 380009}
\address{$^3$IIT Gandhinagar, Chandkheda, Ahmedabad, India-382424}
\ead{nijil@prl.res.in}

\author{Aadhi A}
\address{$^1$Physical Research laboratory, Navarangpura, Ahmedabad, India - 380009}
\ead{aadhi@prl.res.in}

\author{Ali Anwar}
\address{$^1$Physical Research laboratory, Navarangpura, Ahmedabad, India - 380009}
\ead{alianwar@prl.res.in}

\author{R.P. Singh}
\address{$^1$Physical Research laboratory, Navarangpura, Ahmedabad, India - 380009}
\ead{rpsingh@prl.res.in}

\begin{abstract}
We generate the non-separable state of polarization and orbital angular momentum (OAM) using a laser beam. The generated state undergoes a cyclic polarization evolution which introduces a Pancharatnam geometric phase to the polarization state and in turn a relative phase in the non-separable state. We experimentally study the violation of Bell - CHSH inequality for different Pancharatnam phases introduced by various cyclic polarization evolutions with linear and circular states as measurement bases. While measuring in linear bases, the Bell-CHSH parameter oscillates with Pancharatnam phase. One can overcome this dependence by introducing a relative phase in one of the projecting state. However for measurement in circular bases, the Pancharatnam phase does not affect the  Bell-CHSH violation. 

\end{abstract}

%Uncomment for PACS numbers title message
\pacs{42.25.Ja, 41.85.-p, 03.65.Ud, 42.50.Tx}
% Keywords required only for MST, PB, PMB, PM, JOA, JOB? 
\vspace{2pc}
\noindent{\it Keywords}: Orbital angular momentum, non-separable states, Pancharatnam phase
% Uncomment for Submitted to journal title message
%\submitto{\JPA}
% Comment out if separate title page not required
%\maketitle

\section{Introduction}

Non-separability in classical light fields has been studied recently in the context of quantum information and entanglement\cite{spreeuw, simon, ghose, Luis2009, aiello}. This non-separability is analogous to intra-system entanglement involving different degrees of freedom of a light beam.  However, the term "classical entanglement" has received some serious criticism recently \cite{Karimi1172}. The non-separability can exist in between continuous or discrete variables, for which the classical light beams are shown to violate the corresponding form of the Bell's inequality \cite{gsa,shashi, borges, karimi2010}. One can construct classical equivalent of many exotic quantum states using polarization and spatial modes of light. These non separable states are used to mimic many quantum protocols \cite{Hashemi, Perez, goyal}. Apart from the basic interest of demonstrating quantum protocols using classical light beams, they find applications in various fields\cite{toppel,DAmbrosio2013, kagalwala, milione}. These non-separable states are shown to be robust under scattering and used to show the phase recovery of a beam after scattering \cite{Perumangatt, salla}. 

When light fields undergo a cyclic polarization evolution, they acquire a geometric phase which is known as Pancharatnam phase \cite{pancharatnam}. The phase acquired by the light depends on the path taken by the polarization state upon its evolution on the Poincar\'e sphere. The geometric phase is generalized to any quantum system under cyclic evolution by a time dependent Hamiltonian and called as the Berry phase\cite{berry}.  %Here, one of the subsystems will undergo a cyclic evolution which induces a geometric phase on the same subsystem leading to a relative phase in the entangled state which changes its properties such as violation of Bell's inequality.
Entangled states, when generated experimentally, may possess a relative phase due to the phase delays in the generating process. One can nullify this relative phase by introducing a geometric phase in one of the subsystem. In such cases, we need to see the Bell violation, as a measure of entanglement, to optimize the state corresponding to different relative phases. %Thus, it is important to study the effect of geometric phase on the violation of Bell's inequality in the context of different measurement bases.
The effect of Berry phase in entangled systems and their violation of Bell's inequality were studied for spin-$\frac{1}{2}$ particles\cite{bertlmann, Sponar}.  The measurement in two degrees of freedom of non-separable state , be it classical or quantum, is very important for various protocols. Any relative phase will change the measurement outcome which will affect the efficiency of the protocol. Thus optimizing the measurements in two degrees of freedom, when the state acquires a relative phase is very important for efficient implementation of such protocols. Here, we introduce Pancharatnam geometric phase to the non-separable state in a controlled way and show that a proper selection of measurement basis can make the Bell measurement phase independent.

In this article, we generate a classical non-separable Bell-like state of polarization and orbital angular momentum (OAM) of light using a polarizing Sagnac interferometer. We demonstrate the presence of non-separability by the violation of Bell's inequality for the generated state. Next, we study the effect of Pancharatnam phase introduced by the polarization subsystem, through its cyclic evolution on the Poincar\'e sphere, on the violation of Bell's inequality. %We introduce a geometric phase in one of the subsystem, polarization, by evolving it on a cyclic path on the surface of the Poincar\'e sphere. The Pancharatnam phase introduced can be varied continuously by choosing different path for the polarization evolution. We study the effect of this Pancharatnam phase on the violation of Bell's inequality. 
The maximum violation $B_{MAX}$ varies sinusoidally  according to the Pancharatnam phase when maximized over the set of linear  bases. We experimentally show that the Bell parameter can obtain its maximum value of $2\sqrt{2}$ irrespective of the Pancharatnam phase if we introduce a corresponding relative phase in the projecting basis. We also investigate the effect of Pancharatnam phase on the spatially varying polarization structure of these non-separable beams. The result give insight to the measurement optimization of Bell CHSH inequality for a non-separable state under different relative phase.

 In section \ref{sc.2}, we give a theoretical background for the problem. We are using Dirac notation for the representation of states. However it is to be noted that the state correspond to classical electromagnetic fields and also we have omitted the normalization by total intensity for the convenience of representation. In section \ref{sc.3} we describe the experimental details for the generation, evolution and measurement of the non-separable state of light. We also give the holograms for the OAM measurements. The results and discussion are given in section \ref{sc.4}. Along with the projective measurement results, we give the Stokes polarimetric images of the beam for different Pancharatnam phase which can give a physical explanation for the variation of $B_{MAX}$ and the bases adjustments. We finally conclude in section \ref{sc.5}.     
\section{Theoretical Background}\label{sc.2}
We start with a maximally non-separable Bell-like state of polarization and OAM that can be written as  
 \begin{equation}
  \vert \psi\rangle = \frac{1}{\sqrt{2}}\left(\vert H\rangle \vert l\rangle + \vert V \rangle \vert -l\rangle\right)\label{1}
\end{equation}
 where $\vert H\rangle, \vert V \rangle $ and $\vert l\rangle, \vert -l \rangle $ are basis vectors for 2D complex vector spaces of polarization and OAM. We define the set of linear bases corresponding to polarization and OAM as
 \begin{eqnarray}
 \nonumber \vert\theta\rangle &=& \textrm{cos}(\theta) \vert H\rangle+ \textrm{sin}(\theta)\vert V\rangle; \vert\theta_{\perp}\rangle = -\textrm{sin}(\theta) \vert H\rangle+ \textrm{cos}(\theta)\vert V\rangle; \\
  \vert\chi\rangle &=& \textrm{cos}(\chi) \vert l\rangle+ \textrm{sin}(\chi)\vert -l\rangle;\vert\chi_{\perp}\rangle = -\textrm{sin}(\chi) \vert l\rangle+ \textrm{cos}(\chi)\vert -l\rangle;\label{3}
 \end{eqnarray}
 
\begin{figure}[h]
\centering
\includegraphics[scale=.37]{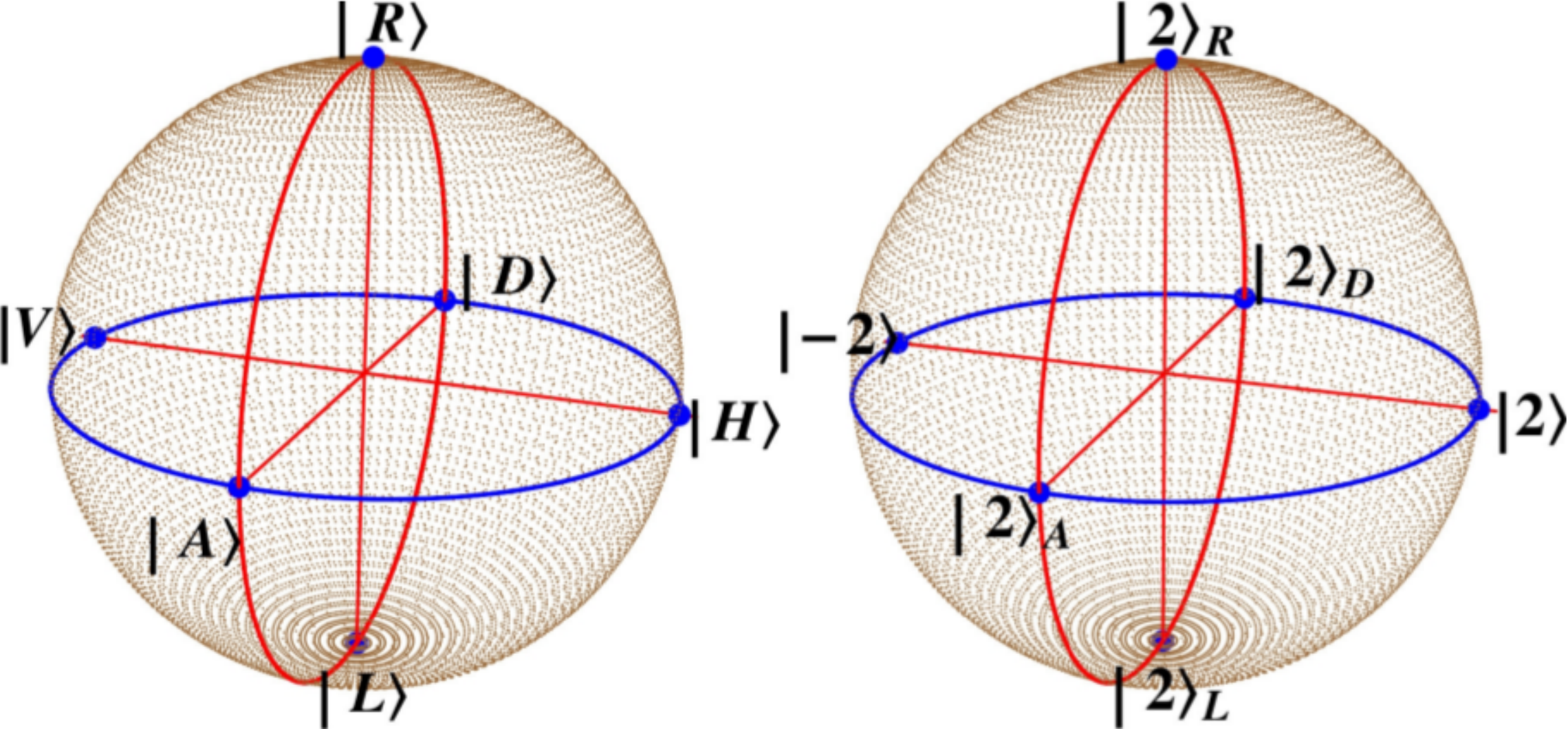}
\caption{ (Color online) Linear (blue circle) and circular (red circles) bases of polarization (left) and OAM(right) states. H, V, D, A, R and L represents the horizontal, verticlal, diagonal, antidiagonal, right circular and left circular polarization. $\vert 2\rangle$ and $\vert -2\rangle$ are OAM states corresponding to the topological charge +2 and -2 respectively. Also $\vert 2\rangle_D = \frac{\vert 2\rangle+\vert -2\rangle}{2}, \vert 2\rangle_A = \frac{\vert 2\rangle-\vert -2\rangle}{2}, \vert 2\rangle_R = \frac{\vert 2\rangle+i\vert -2\rangle}{2} $ and $ \vert 2\rangle_L = \frac{\vert 2\rangle-i\vert -2\rangle}{2}$. }\label{fg.1}
\end{figure}

  These bases are represented geometrically as blue circles on the Poincar\'e sphere of polarization and that of OAM as given in Fig. \ref{fg.1}. The Bell-CHSH inequality is defined as
\begin{equation}
B(\theta,\theta ',\chi,\chi ') = \vert E(\theta,\chi)-E(\theta,\chi ')+ E(\theta ',\chi)+E(\theta ',\chi ')\vert \leq 2 \label{4}
\end{equation}
where 
\begin{equation}
\nonumber E(\theta,\chi) =\frac{C(\theta,\chi)+C(\theta_{\perp},\chi_{\perp})-C(\theta_{\perp},\chi)-C(\theta,\chi_{\perp})}{C(\theta,\chi)+C(\theta_{\perp},\chi_{\perp})+C(\theta_{\perp},\chi)+C(\theta,\chi_{\perp})}. \label{5}
\end{equation}
$C(\theta,\chi)$ is the probability amplitude of a state for being in $\vert\theta\rangle\vert\chi\rangle$. Projecting the state given in Eq. \ref{1} to a general state $\vert\theta\rangle\vert\chi\rangle$ in the linear bases of polarization and OAM we get
 \begin{equation}
C(\theta,\chi) = \vert \langle\theta\vert\langle\chi\vert\psi\rangle\vert^2 \propto \textrm{cos}^2(\theta-\chi)\label{2}
 \end{equation}

Using above expression in Eq. \ref{4}, one can obtain $B_{max} =2\sqrt{2}$ at $(\theta = 0^{\circ},\theta '=45^{\circ},\chi=22.5^{\circ},\chi '=67.5^{\circ})$. Note that the maximization is done over states in the linear bases of polarization and OAM by changing $\theta$ and $\chi$ respectively.  

Now, consider a cyclic evolution of polarization for the state given in Eq. \ref{1} which can be done by action of half wave plate (H) oriented at $45^o$ with the horizontal, quarter wave plate (Q) oriented at $\phi '$ and another half wave plate at $45^{\circ}$.  Here, the two orthogonal polarization states evolve as $H \rightarrow e^{2i\phi'} H $ and $V \rightarrow e^{-2i\phi'} V $. Thus the state becomes

 \begin{equation}
  \vert \psi'\rangle = \frac{1}{\sqrt{2}}\left(\vert H\rangle \vert +l\rangle + e^{-i\phi} \vert V \rangle \vert -l\rangle\right)\label{6}
\end{equation}
 where $\phi = 4\phi'$. The joint measurement probability becomes
 \begin{equation}
 \nonumber C(\theta,\chi,\phi)=  \textrm{cos}^2(\theta)\textrm{sin}^2(\chi)+\textrm{sin}(2\theta)\textrm{sin}(2\chi)\textrm{cos}(\phi)+\textrm{cos}^2(\chi)\textrm{sin}^2(\theta) \label{7}
 \end{equation}
which is a function of $\phi$ also. Thus the  $B_{MAX}$ will also be a function of $\phi$ along with $\theta,\theta', \chi$ and $\chi'$. Changing $\phi$ will affect the angles corresponding $B_{MAX}$ and its value. For $B_{MAX}$ with $\theta = 0^{\circ}, \theta'=90^{\circ}$, $\chi$ and $\chi'$ are varied as
 \begin{equation}
 \chi= \frac{1}{2}\textrm{arctan}(\textrm{cos}\ \phi); \ \chi'=\frac{\pi}{2}-\chi\label{8}.
\end{equation}   
 The $B_{MAX}$ varies periodically with the relative phase $\phi$. 
 \begin{figure*}[h]
\begin{center}
\includegraphics[scale=.7]{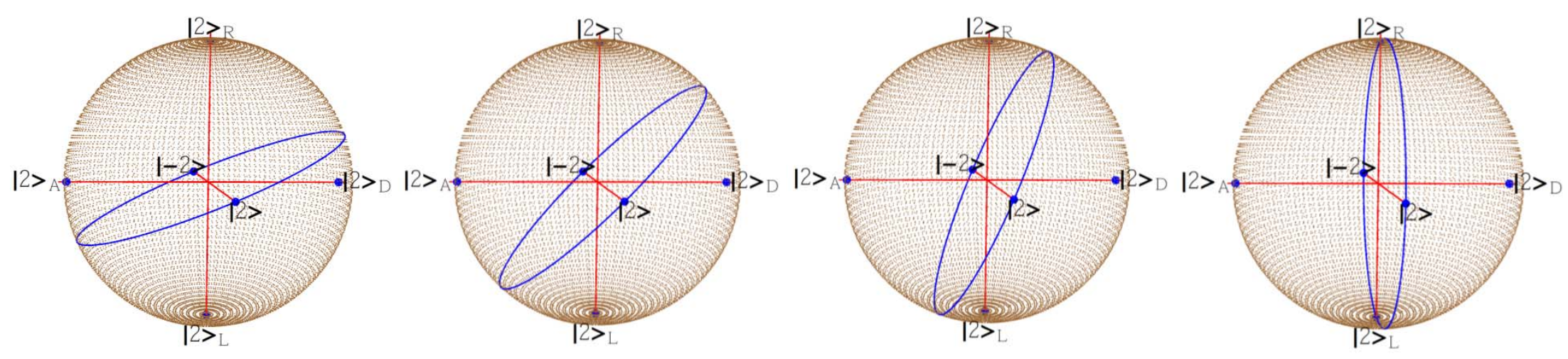} 
\caption{(Color online) Choice of measurment bases for OAM in order to obtain maximum violation for Bell-CHSH parameter for $\phi = 22.5^{\circ}, 45^{\circ}, 67.5^{\circ}$ and $90^{\circ} $ }\label{fg.2}
\end{center}
\end{figure*}
To get back the maximum Bell violation, one need to use different bases for the maximization of the Bell parameter. We redefine the OAM projecting state as
\begin{equation}
\vert\chi'\rangle  = \textrm{cos}(\chi) \vert l\rangle+ e^{-i\phi}  \textrm{sin}(\chi)\vert -l\rangle\label{9}
\end{equation}
that gives
\begin{equation}
C(\theta,\chi')\propto \vert \langle\theta\vert\langle\chi'\vert\psi'\rangle\vert^2=\textrm{cos}^2(\theta-\chi)\label{10}
 \end{equation}
One can obtain the same by introducing a relative phase in polarization state $\vert\theta\rangle$ too. Now the Bell-CHSH parameter $B_{MAX}$ is independent of the relative phase $\phi$. The different OAM measurement bases  for different $\phi$ are given in  Fig. \ref{fg.2}.

Next we check the Bell-CHSH parameter by projecting the state $\vert\psi'\rangle$ in circular basis of polarization and OAM. The states are given as 
\begin{equation}
\vert\theta\rangle= e^{-i\theta}\vert H\rangle+e^{i\theta} \vert V\rangle; \  \vert\chi\rangle= e^{i\chi}\vert l\rangle+e^{-i\chi} \vert -l\rangle.\label{11}
\end{equation} 
These measurement bases are given as red circles in Fig.\ref{fg.1}. Measuring the state $\vert\psi\rangle$ will result the same outcome as given in Eq. \ref{2}.  Now the joint detection probability for the projection of the state $\vert\psi'\rangle$ to state $\vert\theta\rangle\vert\chi\rangle$ is given as
\begin{equation}
\vert\langle\theta\vert\chi\vert\psi'\rangle\vert^2= \textrm{cos}^2(\theta-\chi-\frac{\phi}{2})
\end{equation}
Thus with $\theta=0^{\circ}, \theta'=45^{\circ}, \chi =22.5^{\circ} +\frac{\phi}{2}$ and $\chi' =67.5^{\circ} +\frac{\phi}{2}$ we can obtain the Bell-CHSH parameter as $2\sqrt{2}$.

\section{Experimental Setup}\label{sc.3}

\begin{figure}[h]
\begin{center}
\includegraphics[scale=.8]{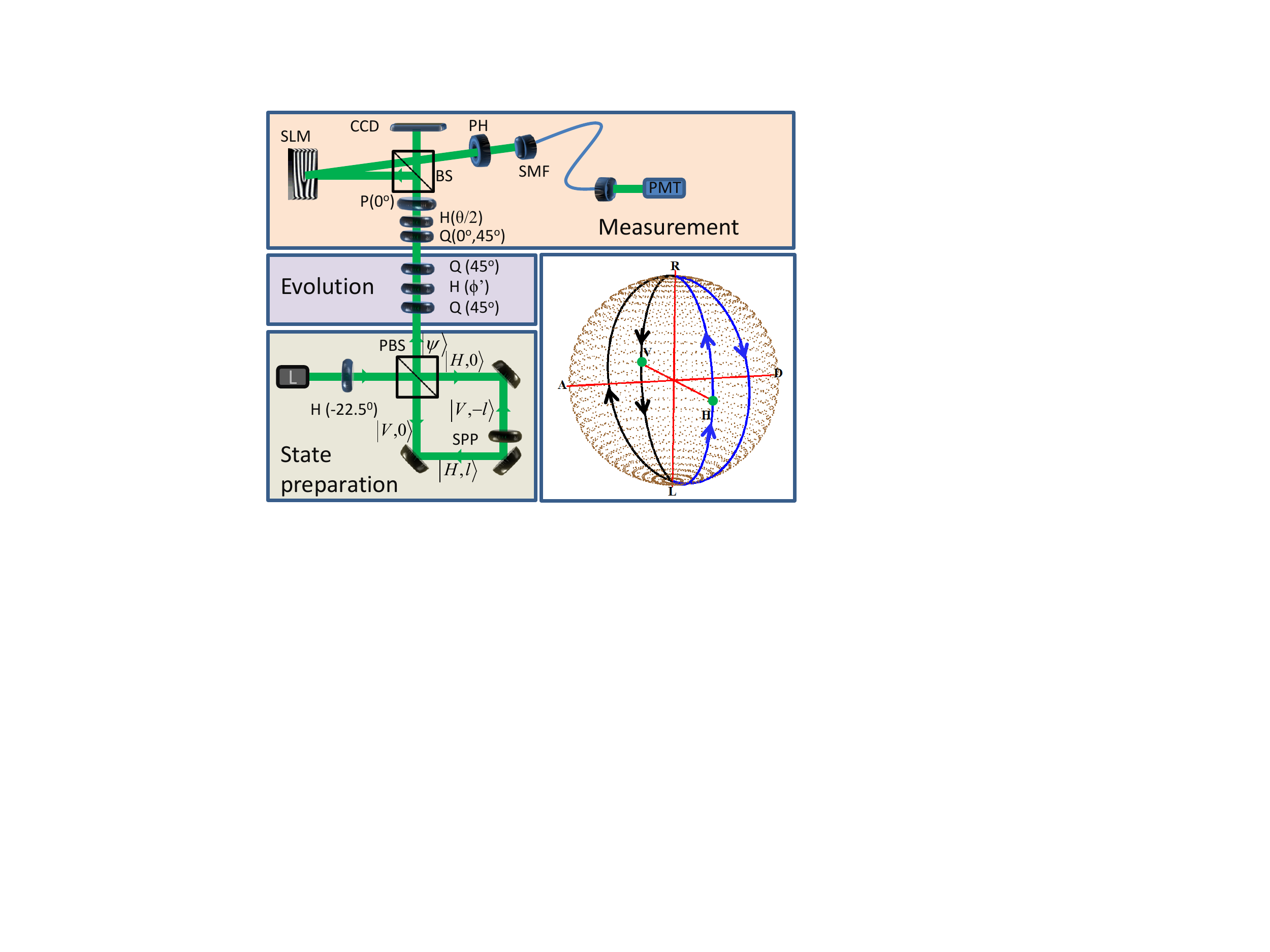} 
\caption{(Color online) Experimental setup for the state preparation, polarization evolution and measurement along with the representation of the polarization evolution in Poincar\'e sphere. L - laser, H - half wave plate, PBS - polarizing beam splitter, SPP - spiral phase plate, Q - quarter wave plate, BS - beam splitter, SLM - spatial light modulator, CCD - charge coupled device (camera), P - polarizer, PH - pin hole, SMF - single mode fiber, PMT - photo multiplier tube.  }\label{fg.3}
\end{center}
\end{figure}

The experimental set up used to generate the non-separable state and to study its properties is shown in Fig. \ref{fg.3}. We have used a diode pumped solid state green laser (Verdi 10) with vertical polarization for our study. The laser beam passes through a half wave plate oriented at $-22.5^o$ with the horizontal that changes the polarization from vertical to diagonal. Then it passes through a polarizing Sagnac interferometer containing a spiral phase plate (SPP) to generate a light beam with non-separable polarization and OAM. Two orthogonally  polarized (H and V) counter propagating Gaussian beams are converted into optical vortices of orders $l$ (for H) and $-l$ (for V) by the SPP designed for order $|l|=2$. These orthogonally polarized and oppositely charged vortices  superpose at the same PBS to form the non-separable state.

\begin{figure}[h]
\begin{center}
\includegraphics[scale=.48]{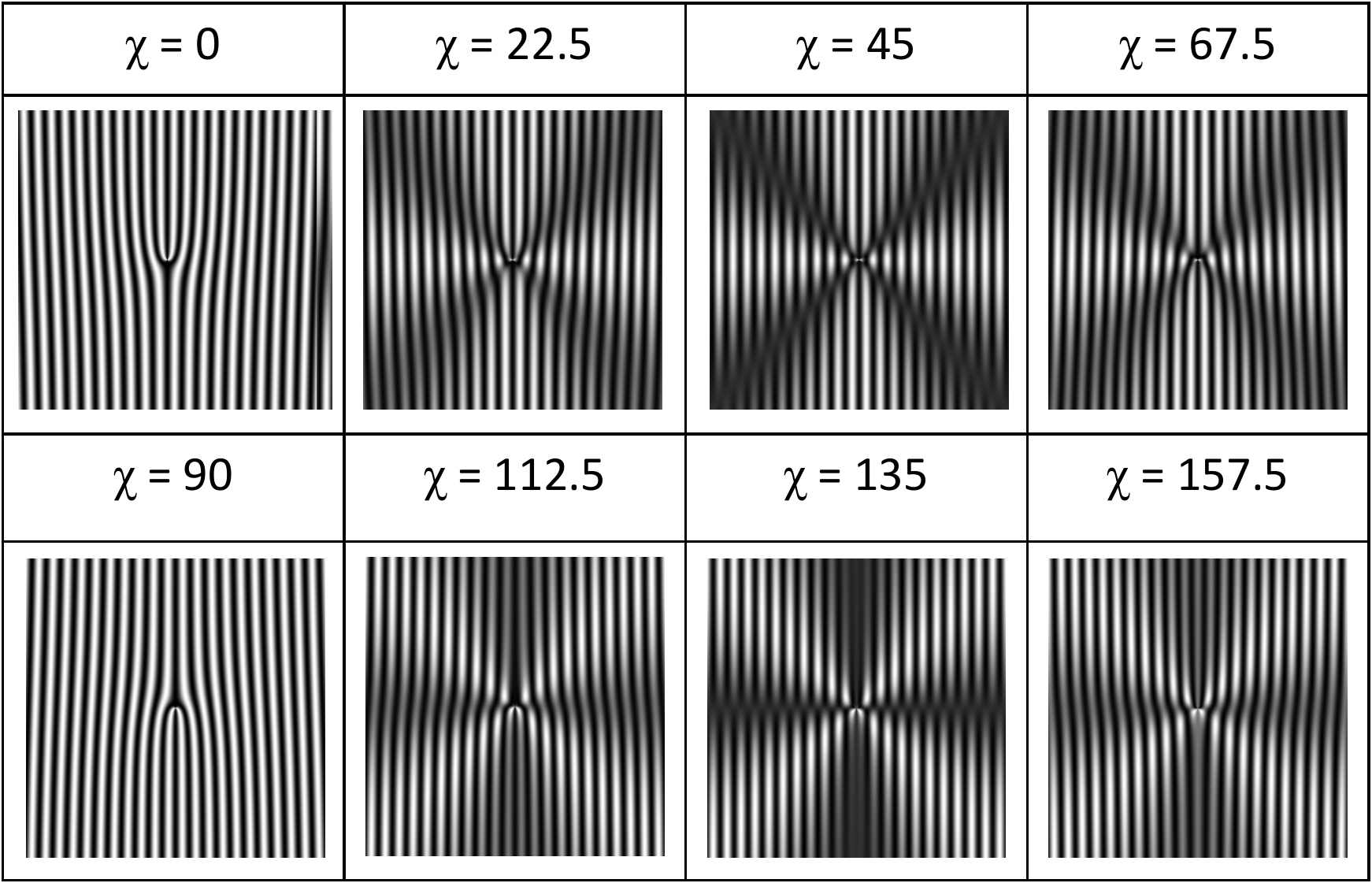} 
\caption{Holograms for different $\chi$ for the measurements of OAM states in linear bases }\label{fg.4}
\end{center}
\end{figure}

For the polarization evolution we have introduced a Simon-Mukunda (SM) gadget \cite{simon1990, Reddy}, a combination of two quarter wave plates (Q) and a half wave plate (H) in Q-H-Q order. A quarter wave plate with fast axis oriented at $45^{\circ}$ with the horizontal convert the horizontal and vertical polarizations into right and left circular polarizations respectively. Now a half wave plate will convert right circular polarization to left circular polarization and vice-versa. The second quarter wave plate at $45^{\circ}$ will convert the circular polarization to the initial linear polarization state.  But the evolution on the Poincar\'e sphere takes different path according to the fast axis orientation angle ($\phi'$) of the half wave plate. The polarization evolution is also given in Fig. \ref{fg.3}.

\begin{figure}[h]
\begin{center}
\includegraphics[scale=.48]{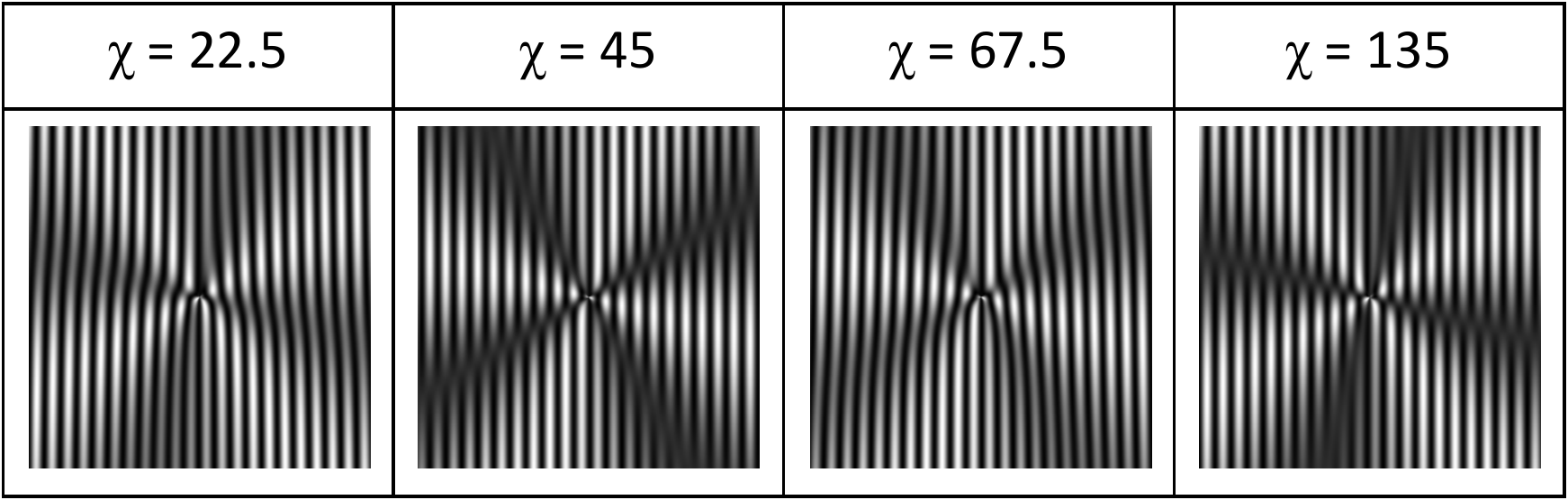} 
\caption{Holograms for optimizing the measurements of OAM states in linear bases with relative phase $\phi=45^{\circ}$ as given in Eq.\ref{9} and $\chi =22.5, 45, 67.5, 135$ }\label{fg.4a}
\end{center}
\end{figure}

The measurement in polarization is done by a quarter wave plate (at 0 or 45 for linear and circular projections), half wave plate (at $\frac{\theta}{2}$) and a polarizer oriented at $0^{\circ}$. For OAM measurements we use a spatial light modulator (SLM) along with a single mode fiber and a photo multiplier tube for the detection. The holograms of the SLM is made in such a way that it converts the particular OAM state $\vert\chi\rangle$ into Gaussian which is coupled to the single mode fiber (SMF). The Holograms for projection in liner OAM bases for different value of $\chi$ are given in Fig. \ref{fg.4}. %The circular projection of OAM at $\chi=0$,  correspond to the $\chi =45^{\circ}$ hologram shown in Fig. \ref{fg.4}. For other values of $\chi$ the same hologram is used with a rotation $\frac{\chi}{2}$. 
The hologram for $\chi=0$ in circular bases is equivalent to the linear basis hologram corresponding to $\chi = 45^{\circ}$ given in Fig. \ref{fg.4}. For other values of $\chi$, the hologram is rotated at an angle $\frac{\chi}{2}$. Fig. \ref{fg.4a} gives the holograms for the optimized measurement of Bell parameter as given in Eq.\ref{9} with $\phi=45$ and for different $\chi$. 

To study the spatially varying  polarization structure and its evolution with the introduced geometric phase, we have carried out the Stokes polarimetric imaging of the beam. For this we image the beam after the polarizer (P) using a CCD camera for the projection to three sets of orthogonal polarization states. Spatially varying Stokes parameters are measured from the images corresponding to different polarization projections.

\section{Results and discussion}\label{sc.4}
\begin{figure}[h]
\begin{center}
\includegraphics[scale=.8]{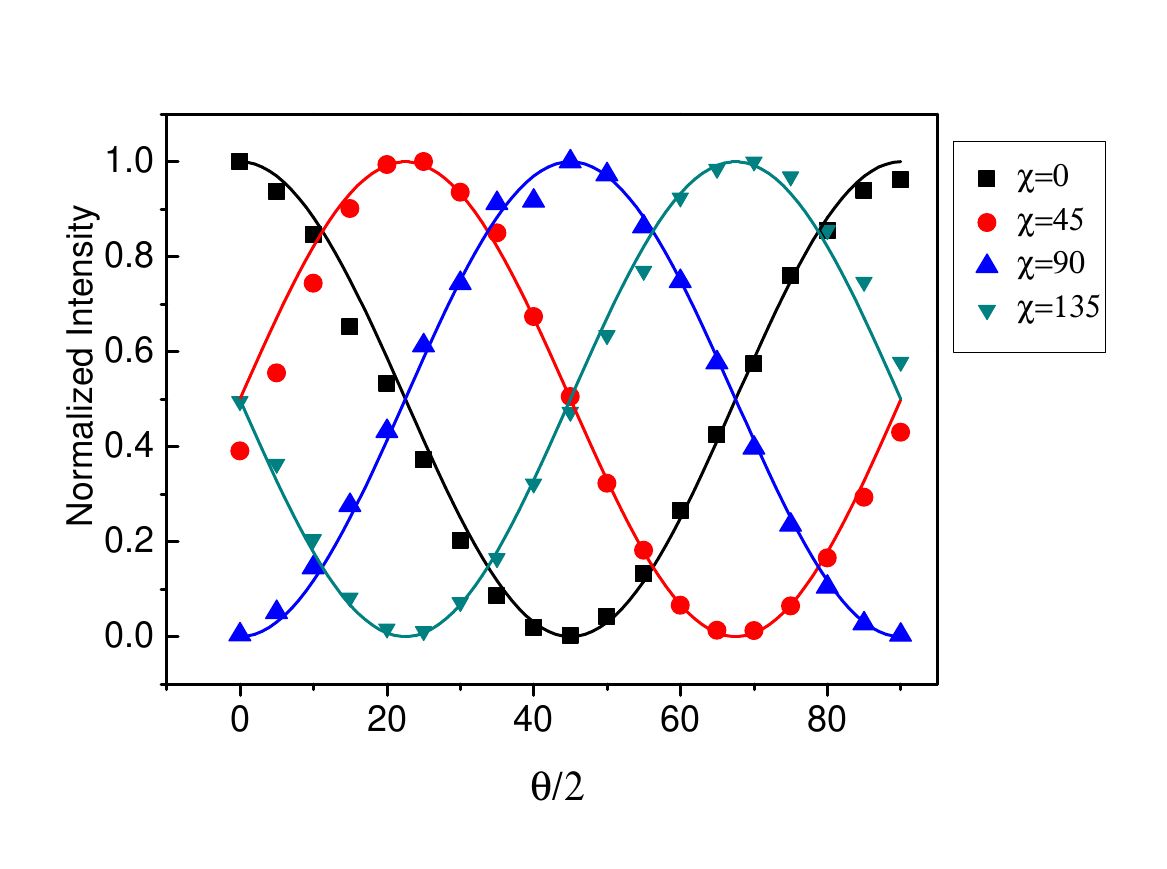} 
\caption{(Color online)Joint polarization-OAM measurement results for a non-seperable state given in Eq. \ref{1}. Measurements are done in linear bases }\label{fg.5}
\end{center}
\end{figure}

Measurements on a light beam with non-separable state of polarization and OAM give raise to contexual results. The measurements are similar to two photon correlation experiments in entangled photon pairs. Instead of measuring one quantity, say polarization, of two spatially separated photons, here we are performing a measurement in two independent degrees of freedom, namely polarization and OAM, of the same beam. The measurement in polarization affects the measurement outcome in OAM. We measure the power coupled to the single mode fibre after the polarization projection by the wave plates and the polarizer and OAM projection by the SLM. We vary the $\theta$ by changing HWP orientation for different values of OAM projection angle $\chi$. The curves for the state given in Eq. \ref{1}, are given in Fig. \ref{fg.5}.The theoretical curves which follow the Eq. \ref{2} are also given for comparison.
\begin{figure}[h]
\begin{center}
\includegraphics[scale=.35]{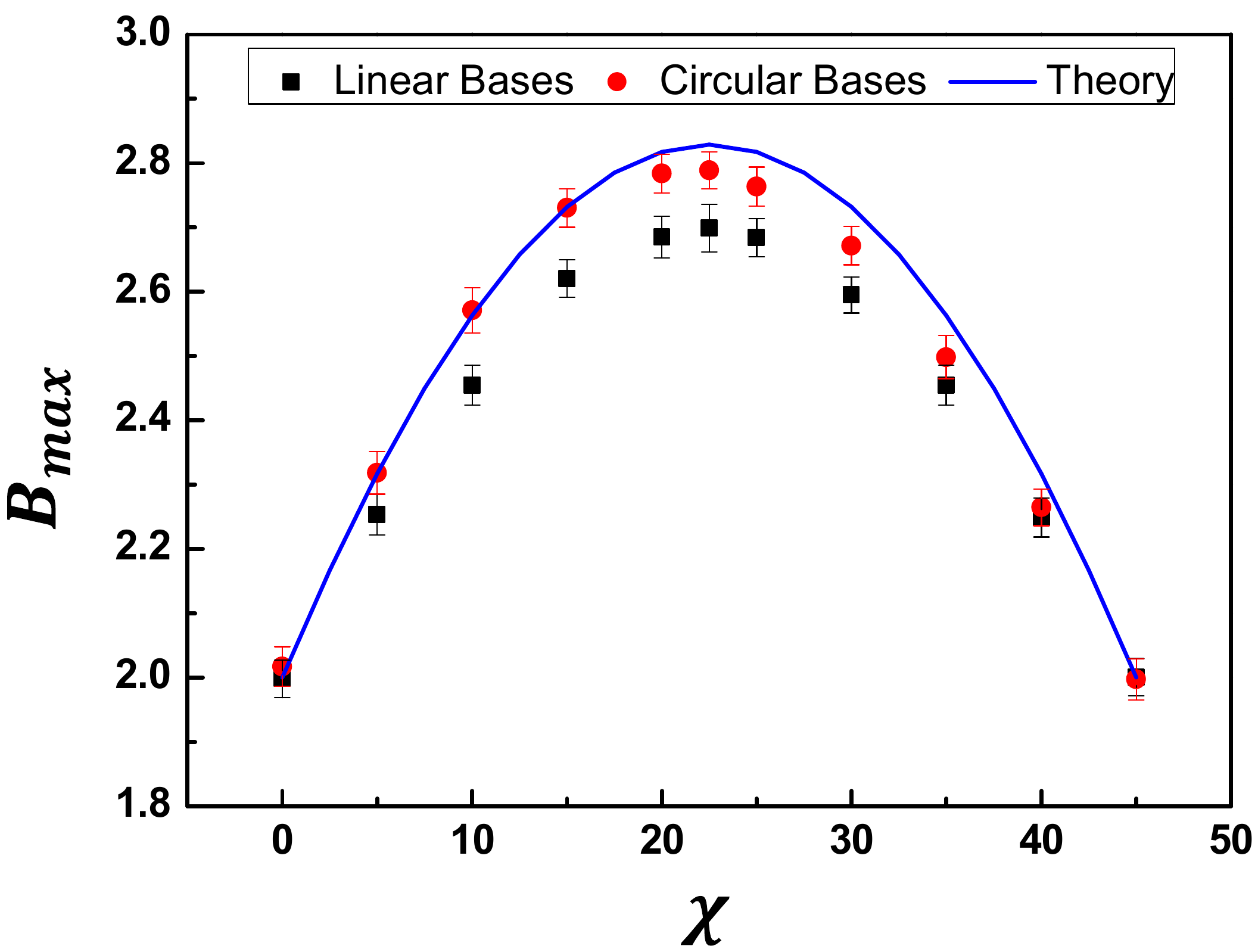}
\caption{(Color online) Bell-CHSH parameter for different measurement angle $\chi$.  }\label{fg.6}
\end{center}
\end{figure}

 To find the optimum angles for the maximum violation of Bell-CHSH inequality, we vary the projecting angle from $0^{\circ}$ to $45$ fixing $\chi' = \chi+45^{\circ}, \theta = 0^{\circ}$ and $\theta '=45^{\circ}$ . For the generated non-separable state, we have  carried out the measurements in linear (black squares) and circular bases (red circles) and the results are given in Fig. \ref{fg.6}. Theoretically expected curve is also given in comparison. With $(\theta = 0^{\circ},\theta '=45^{\circ},\chi=22.5^{\circ},\chi '=67.5)$ we obtain $B_{MAX} = 2.69\pm0.036$ corresponding to $\phi =0^{\circ}$ when measured in linear bases. Measurement in circular bases with the angles mentioned yields a value of  $B_{MAX} = 2.79\pm0.029$. The violation of Bell-CHSH inequality indicates the presence of non-separability of polarization and OAM present in the beam. This accounts for the contexuality  in measuring these two properties of light. The result of OAM measurement depends on the polarization measurement settings. The imperfections in the linear bases projection of OAM using SLM and single mode fiber results in the lesser violation of Bell CHSH inequality. When projecting the input OAM state (after the polarization projections) by the holograms given in Fig. \ref{fg.4}, the center of the Gaussian mode at the fiber coupler slightly get shifted for different projections which affects the coupling to the single mode fiber.  In the case of projections to circular bases, the centers of the projected Gaussian mode are comparatively stable due to the circular symmetry of the different holograms.  
 
\begin{figure}[h]
\begin{center}
\includegraphics[scale=.3]{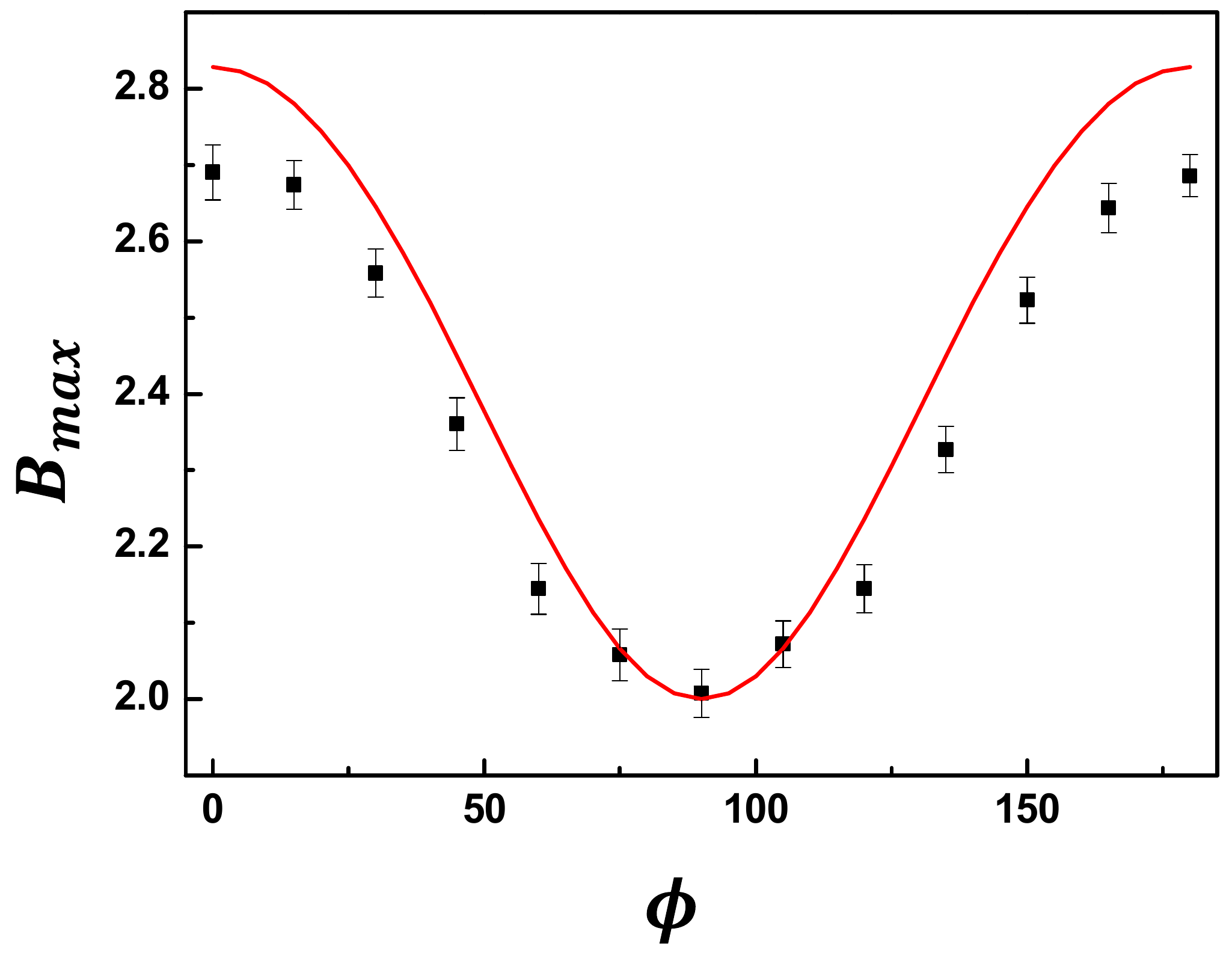} 
\caption{(Color online)Variation of $B_{MAX}$ with the relative phase when maximized over the linear bases }\label{fg.7}
\end{center}
\end{figure}
\begin{figure}[h]
\begin{center}
\includegraphics[scale=.3]{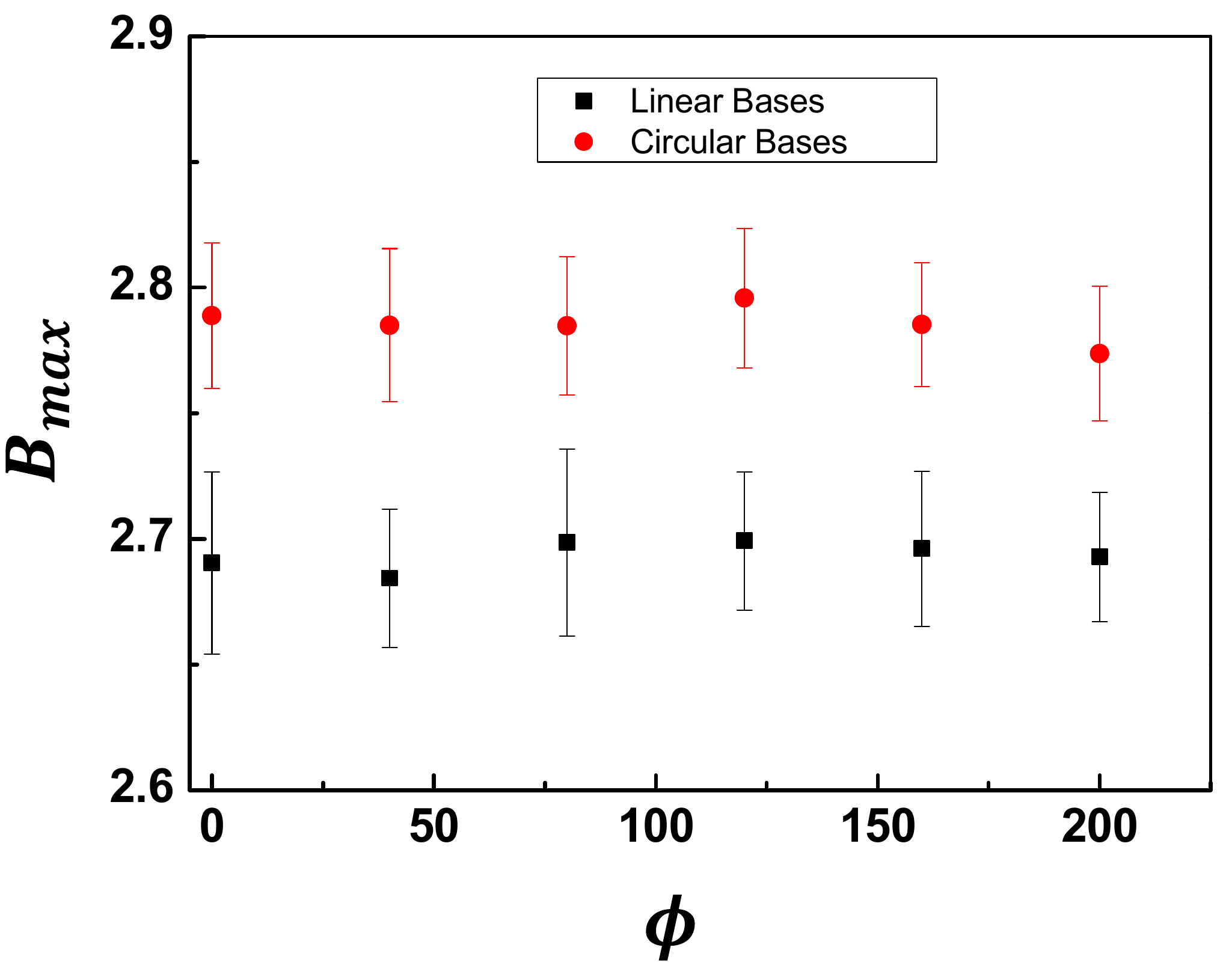} 
\caption{(Color online)Measured values of phase independent $B_{MAX}$ by introducing phase compensation in linear bases (black squares) and changing $\chi$ in circular bases }\label{fg.8}
\end{center}
\end{figure}

 We have changed the relative phase $\phi$ using the SM gadget and obtained $B_{MAX}$ using conditions given in Eq. \ref{8}. Experimental curve for the $B_{MAX}$ with the phase $\phi$ is given in Fig.\ref{fg.7}. At $\phi = 90$ the value of $B_{MAX}$ drops to zero, and there is no violation of Bell inequality. However, the state given in Eq. \ref{6} is always maximally non-separable/ entangled as the other entanglement measures like concurrence or von-Newman entropy are independent of $\phi$.

We have carried out the Bell parameter measurement with the introduction of relative phase in $\vert\chi\rangle$ to transform it to $\vert\chi\rangle$ as given in Eq. \ref{9}. The $\phi$ independent values of $B_{MAX}$ for these measurements are given in Fig. \ref{fg.8}. We also measure in circular basis for which projecting states are described by Eq. \ref{11}.

\begin{figure}[h]
\begin{center}
\includegraphics[scale=.32]{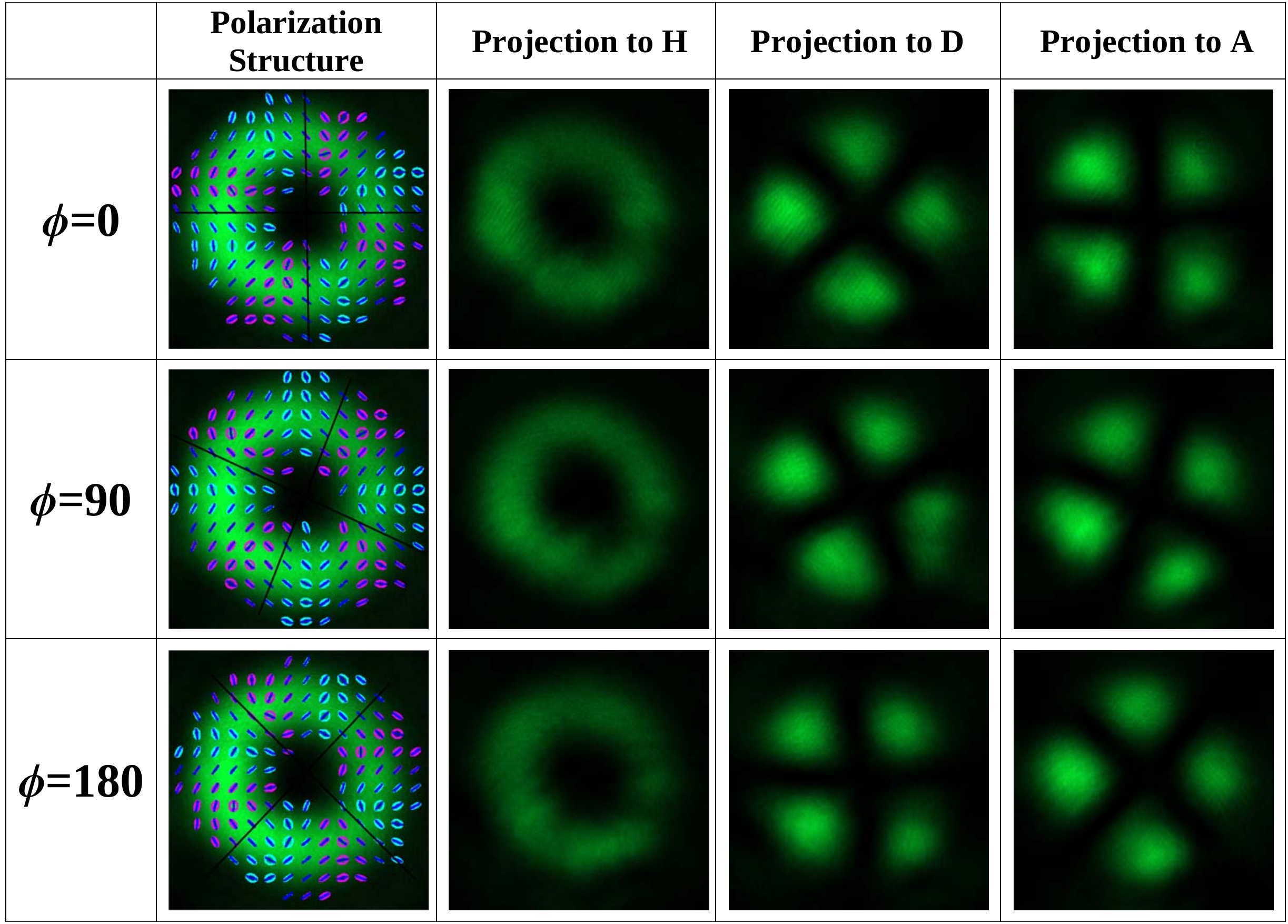} 
\caption{(Color online)Polarization structure and the intensity profile corresponding to different polarization projections for different relative phase $\phi$}\label{fg.9}
\end{center}
\end{figure}

With the change in measurement angles as mentioned above Bell - CHSH parameter is found to be constant with the relative phase $\phi$. The results are given in Fig. \ref{fg.8}. Here we don't have to change the measurement basis, as we have done in the linear case, to maximize $B$. So, when projecting in circular basis, one can easily compensate the effect of Pancharatnam phase in the Bell-CHSH inequality measurement.

 We have analyzed the spatially varying polarization structure with the cyclic polarization evolution. The results are given in Fig. \ref{fg.9}. The magenta and cyan circles show right and left circular polarizations. Two black lines are drawn corresponding to the diagonal polarization. It is found that the total polarization structure rotates with the relative phase introduced. We also give the images corresponding to different polarization projections for different $\phi$. One can see that the images corresponding to diagonal and anti-diagonal projections are rotated with the relative phase $\phi$. However, it doesn't affect the mode structure corresponding to horizontal or vertical projection. For the projection in any other linear state, the mode rotates with the relative phase. Thus a hologram that is supposed to convert the diagonal OAM state $\frac{1}{\sqrt{2}} (\vert l\rangle +\vert -l\rangle ) $ to Gaussian, will not effectively convert the state $\frac{1}{\sqrt{2}} (\vert l\rangle +e^{i\phi} \vert -l\rangle ) $. By introducing a phase in the OAM projecting bases as given in Eq. \ref{9}, which gives a rotation for the hologram corresponding to the projection onto $\chi$ other than $0^{\circ},90^{\circ}$ , one can compensate this effect and achieve the maximum violation of Bell CHSH inequality as given in Fig. \ref{fg.8}. The optimized holograms are given in Fig. \ref{fg.4a} for $\phi =45$.  \\

\section{conclusions}\label{sc.5}
In conclusion, we have studied the effect of Pancharatnam geometric phase in a non-separable state of polarization and OAM. The non-separability is confirmed by the violation of Bell-CHSH inequality. The geometric phase introduced in the polarization subsystem induces a relative phase in the Bell like state of OAM and polarization. The maximum value of the Bell parameter $B_{MAX}$, maximized over the measurement angles, varies sinusoidally according to the relative phase. We obtain a constant $B_{MAX}$ for different geometric phase by introducing a relative phase in the projected OAM state. We also show that the Bell CHSH inequality measurement in circular bases can remove the phase dependence of the $B_{MAX}$ by shifting the measurement angle. We have analyzed the polarization structure of the non-separable state for different Pancharatnam phases which gives a rotation to it. This physically explain the effect of Pancharatnam phase in the joint measurement of polarization and OAM.  
\section{acknowledgment}
Authors like to acknowledge Prof. Partha Ghose, SRFTI, Kolkatta for the fruitful discussions. Also we would like to acknowledge Dr. Vijay Kumar, PDF, Physical Research Laboratory for the polarization Stokes images.


\section*{References}
\begin{thebibliography}{10}

\bibitem{spreeuw}
Robert J.~C. Spreeuw.
\newblock A classical analogy of entanglement.
\newblock {\em Found. Phys.}, 28(3):361--374, 1998.

\bibitem{simon}
B.~N. Simon, S.~Simon, F.~Gori, M.~Santarsiero, R.~Borghi, N.~Mukunda, and
  R.~Simon.
\newblock Nonquantum entanglement resolves a basic issue in polarization
  optics.
\newblock {\em Phys. Rev. Lett.}, 104:023901, Jan 2010.

\bibitem{ghose}
Partha Ghose and Anirban Mukherjee.
\newblock Entanglement in classical optics.
\newblock {\em Rev. Theor. Sci.}, 2(4):274--288, 2014.

\bibitem{Luis2009}
Alfredo Luis.
\newblock Coherence, polarization, and entanglement for classical light fields.
\newblock {\em Optics Communications}, 282(18):3665 -- 3670, 2009.

\bibitem{aiello}
Andrea Aiello, Falk T{\"o}ppel, Christoph Marquardt, Elisabeth Giacobino, and
  Gerd Leuchs.
\newblock Quantum−like nonseparable structures in optical beams.
\newblock {\em New Journal of Physics}, 17(4):043024, 2015.

\bibitem{Karimi1172}
Ebrahim Karimi and Robert~W. Boyd.
\newblock Classical entanglement?
\newblock {\em Science}, 350(6265):1172--1173, 2015.

\bibitem{gsa}
Priyanka Chowdhury, A.~S. Majumdar, and G.~S. Agarwal.
\newblock Nonlocal continuous-variable correlations and violation of bell's
  inequality for light beams with topological singularities.
\newblock {\em Phys. Rev. A}, 88:013830, Jul 2013.

\bibitem{shashi}
Shashi Prabhakar, Salla~Gangi Reddy, A.~Aadhi, Chithrabhanu Perumangatt, G.~K.
  Samanta, and R.~P. Singh.
\newblock Violation of bell's inequality for phase-singular beams.
\newblock {\em Phys. Rev. A}, 92:023822, Aug 2015.

\bibitem{borges}
C.~V.~S. Borges, M.~Hor-Meyll, J.~A.~O. Huguenin, and A.~Z. Khoury.
\newblock Bell-like inequality for the spin-orbit separability of a laser beam.
\newblock {\em Phys. Rev. A}, 82:033833, Sep 2010.

\bibitem{karimi2010}
Ebrahim Karimi, Jonathan Leach, Sergei Slussarenko, Bruno Piccirillo, Lorenzo
  Marrucci, Lixiang Chen, Weilong She, Sonja Franke-Arnold, Miles~J. Padgett,
  and Enrico Santamato.
\newblock Spin-orbit hybrid entanglement of photons and quantum contextuality.
\newblock {\em Phys. Rev. A}, 82:022115, Aug 2010.

\bibitem{Hashemi}
Seyed~Mohammad Hashemi~Rafsanjani, Mohammad Mirhosseini, Omar~S. Maga\~na
  Loaiza, and Robert~W. Boyd.
\newblock State transfer based on classical nonseparability.
\newblock {\em Phys. Rev. A}, 92:023827, Aug 2015.

\bibitem{Perez}
Benjamin Perez-Garcia, Melanie McLaren, Sandeep~K. Goyal, Raul~I.
  Hernandez-Aranda, Andrew Forbes, and Thomas Konrad.
\newblock Quantum computation with classical light: Implementation of the
  deutsch–jozsa algorithm.
\newblock {\em Physics Letters A}, 380(22–23):1925 -- 1931, 2016.

\bibitem{goyal}
Sandeep~K. Goyal, Filippus~S. Roux, Andrew Forbes, and Thomas Konrad.
\newblock Implementing quantum walks using orbital angular momentum of
  classical light.
\newblock {\em Phys. Rev. Lett.}, 110:263602, Jun 2013.

\bibitem{toppel}
Falk T{\"o}ppel, Andrea Aiello, Christoph Marquardt, Elisabeth Giacobino, and
  Gerd Leuchs.
\newblock Classical entanglement in polarization metrology.
\newblock {\em New Journal of Physics}, 16(7):073019, 2014.

\bibitem{DAmbrosio2013}
Vincenzo D'Ambrosio, Nicol{\`{o}} Spagnolo, Lorenzo~Del Re, Sergei Slussarenko,
  Ying Li, Leong~Chuan Kwek, Lorenzo Marrucci, Stephen~P. Walborn, Leandro
  Aolita, and Fabio Sciarrino.
\newblock Photonic polarization gears for ultra-sensitive angular measurements.
\newblock {\em Nature Communications}, 4:022115, sep 2013.

\bibitem{kagalwala}
Kumel~H Kagalwala, Giovanni Di~Giuseppe, Ayman~F Abouraddy, and Bahaa~EA Saleh.
\newblock Bell's measure in classical optical coherence.
\newblock {\em Nature Photon.}, 7(1):72--78, 2013.

\bibitem{milione}
Giovanni Milione, Thien~An Nguyen, Jonathan Leach, Daniel~A Nolan, and Robert~R
  Alfano.
\newblock Using the nonseparability of vector beams to encode information for
  optical communication.
\newblock {\em Opt. Lett.}, 40(21):4887--4890, 2015.

\bibitem{Perumangatt}
Chithrabhanu Perumangatt, Gangi~Reddy Salla, Ali Anwar, A.~Aadhi, Shashi
  Prabhakar, and R.P. Singh.
\newblock Scattering of non-separable states of light.
\newblock {\em Opt. Commun.}, 355:301 -- 305, 2015.

\bibitem{salla}
Gangi~Reddy Salla, Chithrabhanu Perumangattu, Shashi Prabhakar, Ali Anwar, and
  Ravindra~P. Singh.
\newblock Recovering the vorticity of a light beam after scattering.
\newblock {\em Appl. Phys. Lett.}, 107(2):--, 2015.

\bibitem{pancharatnam}
Shivaramakrishnan Pancharatnam.
\newblock Generalized theory of interference and its applications.
\newblock In {\em Proc. Indian Acad. Sci. Sect. A}, volume~44, page 247.
  Springer, 1956.

\bibitem{berry}
Michael~V Berry.
\newblock Quantal phase factors accompanying adiabatic changes.
\newblock In {\em Proc. R. Soc. London, Ser. A}, volume 392, page~45. The Royal
  Society, 1984.

\bibitem{bertlmann}
Reinhold~A. Bertlmann, Katharina Durstberger, Yuji Hasegawa, and Beatrix~C.
  Hiesmayr.
\newblock Berry phase in entangled systems: A proposed experiment with single
  neutrons.
\newblock {\em Phys. Rev. A}, 69:032112, Mar 2004.

\bibitem{Sponar}
S.~Sponar, J.~Klepp, R.~Loidl, S.~Filipp, K.~Durstberger-Rennhofer, R.~A.
  Bertlmann, G.~Badurek, H.~Rauch, and Y.~Hasegawa.
\newblock Geometric phase in entangled systems: A single-neutron interferometer
  experiment.
\newblock {\em Phys. Rev. A}, 81:042113, Apr 2010.

\bibitem{simon1990}
R.~Simon and N.~Mukunda.
\newblock Minimal three-component su(2) gadget for polarization optics.
\newblock {\em Physics Letters A}, 143(4):165 -- 169, 1990.

\bibitem{Reddy}
Salla~Gangi Reddy, Shashi Prabhakar, A.~Aadhi, Ashok Kumar, Megh Shah, R.~P.
  Singh, and R.~Simon.
\newblock Measuring the mueller matrix of an arbitrary optical element with a
  universal su(2) polarization gadget.
\newblock {\em J. Opt. Soc. Am. A}, 31(3):610--615, Mar 2014.

\end{thebibliography}
\end{document}